\documentclass[pra,showpacs,superscriptaddress,amsmath,amssymb]{revtex4}
\usepackage{amsmath}
\usepackage{graphicx}
\usepackage{bm}

\begin{document}

\title{Interference-induced enhancement of field entanglement in a microwave-driven V-type single-atom laser}

\author{Wen-Xing Yang} \email{wenxingyang2@126.com}
\affiliation{Department of Physics, Southeast University, Nanjing
210096, China}\affiliation{Institute of Photonics Technologies,
National Tsing-Hua University, Hsinchu 300, Taiwan}
\author{Ai-Xi Chen}
\affiliation{Department of Applied Physics, School of Basic Science,
East China Jiaotong University, Nanchang 330013, China}
\author{Ting-Ting Zha}
\affiliation{Department of Physics, Southeast University, Nanjing
210096, China}
\author{Yanfeng Bai}
\affiliation{Department of Physics, Southeast University, Nanjing 210096, China}
\author{Ray-Kuang Lee}
\affiliation{Institute of Photonics Technologies, National Tsing-Hua
University, Hsinchu 300, Taiwan}

\date{\today}

\begin{abstract}
We investigate the generation and the evolution of two-mode
continuous-variable (CV) entanglement from system of a
microwave-driven V-type atom in a quantum beat laser. By taking into
account the effects of spontaneously generated quantum interference
between two atomic decay channels, we show that the CV entanglement
with large mean number of photons can be generated in our scheme,
and the property of the filed entanglement can be adjusted by
properly modulating the frequency detuning of the fields. More
interesting, it is found that the entanglement can be significantly
enhanced by the spontaneously generated interference.
\\
\keywords{continuous-variable entanglement; quantum interference; quantum information}
\end{abstract}

\pacs{42.50.Dv, 03.67.Mn}

\maketitle

\section{Introduction}
Quantum entanglement has become a fundamental resource for quantum information science, as it takes on extensive potential in the application of quantum computation and quantum communication \cite{1,2,5,6,7,8,9,10,11,12,13,14}. In particular, because of the
relative simplicity and high efficiency in the generation,
manipulation and detection of optical continuous-variable (CV)
states \cite{15,16,17,18}, CV entanglement can offer an
advantage in quantum information processing \cite{20}, and it has
been an important part of quantum information theory \cite{20}.
Therefore, more and more theoretical and experimental efforts have
been devoted to the generation of CV entanglement
\cite{21,24,25,26,w1,w2,w3}. Meanwhile, on the aspect of theory, Simon
\cite{27} and Duan et al \cite{28} have proposed inseparability
criterions for CV states, separately.

It has been proven to be efficient ways for generating the CV
entangled beams that using of Nondegenerate parametric down
conversion (NPDC) in a crystal \cite{15,29}. Besides, the
preparation of the CV entangled light based on the interaction of
two-mode cavity fields with atoms coherently driven by laser fields
has also been investigated extensively. For example, Li et al.
\cite{30} considered the generation of two-mode entangled states of
the cavity field via the four-wave mixing process, by means of the
interaction of properly driven V-type three-level atoms with two
cavity modes. Subsequently, Tan et al \cite{31} extended the
analysis of \cite{30} and studied the generation and evolution of
entangled light by taking into account the effects of spontaneously
generated interference between two atomic decay channels. In an
earlier study, Qamar et al. \cite{32} proposed a scheme for
generating of two-mode entangled states in a quantum beat laser
\cite{33}. The system consists of a V-type three-level atom
interacting with two modes of the cavity field in a doubly resonant
cavity, and the atom is driven into a coherent superposition of the
upper two levels by a strong classical field. They numerically
studied the property of entanglement for different values of Rabi
frequencies in the presence of cavity losses. And Fang et al
\cite{34} extended the analysis of \cite{32}, investigating the
influence of phase and Rabi frequency of the classical driving
field, cavity loss, and the purity and nonclassicality of the
initial state of the cavity field on the property of the resulting
two-mode entangled state. Moreover, in recent years, more and more
theoretical and experimental efforts \cite{26,35,36,37,38} have been
devoted to the generation of entanglement in macroscopic light based
on the single-atom laser.

Following by the work \cite{32} and \cite{34}, we study the
generating and evolution of two-mode CV entangled states  from
system of V-type atom in a quantum beat laser \cite{33} by taking
into account the effects of spontaneously generated interference. In
our scheme, the two transitions in the V-type atom independently
interact with the two cavity modes and the two upper levels of the
atom are driven by a strong classical field. We show that the CV
entanglement with large mean number of photons can be generated in
our scheme and  by properly modulating the frequency detuning of the
fields, the property of the filed entanglement can be adjusted. More
interesting, it is found that the entanglement can be significantly
enhanced by the spontaneously generated interference, in the given
situation.

\section{model and master equations}
Let us consider the atomic system for the quantum beat laser which
is proposed by Scully and Zubairy \cite{33}. It is consist of a
three-level atom with the V configuration interacting with two
(nondegenerate) cavity modes and the two upper levels of the atom
are driven by a strong classical field. In Fig. 1, we show the the
atomic level scheme.

The atom is pumped at a rate $\gamma_a$ into the level
$\left|a\right\rangle$. Use a strong magnetic field with Rabi
frequency $\Omega$ to drive the transition between
$\left|a\right\rangle$ and $\left|b\right\rangle$, which is
electric-dipole forbidden. While the two nondegenerate cavity modes
of frequencies $\nu_1$ and $\nu_2$ independently interact with the
transitions
$\left|a\right\rangle\leftrightarrow\left|c\right\rangle$ (with
$\omega_{ac}$ resonant frequency) and
$\left|b\right\rangle\leftrightarrow\left|c\right\rangle$ (with
$\omega_{bc}$ resonant frequency), respectively.
$\Delta_1=\omega_{ac}-\nu_1$ is the detuning of the field $g_1$ from
the corresponding atomic transition
$\left|a\right\rangle\leftrightarrow\left|c\right\rangle$.
$\Delta_2=\omega_{bc}-\nu_2$ is the detuning of the field $g_2$ from
the corresponding atomic transition
$\left|b\right\rangle\leftrightarrow\left|c\right\rangle$.For
simplicity, we'll note $\Delta_1=\Delta_2=\Delta$.

\begin{figure}
\includegraphics[width=9cm]{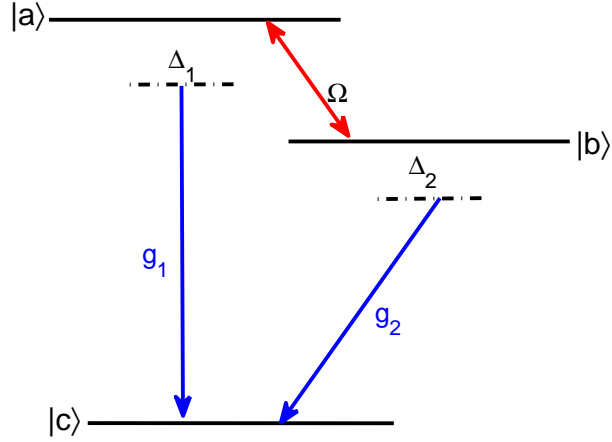}
\caption{\label{Fig1} Schematic diagram of the three-level atom
system in a V configuration. Two (nondegenerate) cavity modes with
coupling constant $g_1$ and $g_2$ interact with the transition
$\left|a\right\rangle\leftrightarrow\left|c\right\rangle$ and
$\left|b\right\rangle\leftrightarrow\left|c\right\rangle$,
respectively, while the atom transition
$\left|a\right\rangle\leftrightarrow\left|b\right\rangle$ is driven
by a strong magnetic field with Rabi frequency $\Omega$. $\Delta_1$
and $\Delta_2$ correspond the frequency detunings.}
\end{figure}

Then, under the dipole and rotating wave approximation, the total
interaction Hamiltonian of our system can be given in the
interaction picture by ($\hbar=1$)
\begin{equation}
\label{eq1} V_I=\Delta\left|a\right\rangle\left\langle
a\right|+\Delta\left|b\right\rangle\left\langle
b\right|+(-\Omega\left|a\right\rangle\left\langle
b\right|+g_1a_1\left|a\right\rangle\left\langle
c\right|+g_2a_2\left|b\right\rangle\left\langle c\right|+H.c.),
\end{equation}
\noindent where the symbol $H.c.$ means the Hermitian conjugate and
we have taken the ground state $\left|c\right\rangle$ as the energy
origin for the sake of simplicity. $\Omega=|\Omega|\exp{(i\phi)}$
describe the Rabi frequencies of the strong classical field and
denote the phase of classical fields with $\phi$, and $g_1$ and
$g_2$ are the atom$-$field coupling constants. $a_j$ ($a^{\dag}_j$)
is the annihilation (creation) operator of the corresponding cavity
modes.

Considering the vacuum damping of the atom and the cavity modes, the
reduced density equations of the cavity fields (taking a trace over
the atom degrees of freedom \cite{39}) can be obtained from the
Hamiltonian (1):
\begin{eqnarray}
\label{eq2}
\dot{\rho}_{f}&=&-iTr_{atom}[V_I,\rho]+Tr_{atom}(L_f\rho+L_a\rho)\nonumber\\
&=&-ig_1[a_1^{\dag},\rho_{ac}]-ig_2[a_2^{\dag},\rho_{bc}]+\sum_{j=1}^{2}\kappa_j[a_j,\rho_f
a_j^{\dag}]+H.c.,
\end{eqnarray}
\noindent with
\begin{eqnarray}
\label{eq3} &&L_f\rho=\sum_{j=1}^{2}\kappa_j[a_j,\rho
a_j^{\dag}]+H.c.,\\
\label{eq4} &&L_a\rho=\gamma_1[s_{ca},\rho
s_{ac}]+\gamma_2[s_{cb},\rho s_{bc}]+\gamma_{12}([s_{ca},\rho
s_{bc}]+[s_{cb},\rho s_{ac}])+H.c.,
\end{eqnarray}
\noindent where $L_f\rho$, $L_a\rho$ are the vacuum damping of the
cavity modes and atom, respectively. $\gamma_1$, and $\gamma_2$ are
the decay rates from the states $\left|a\right\rangle$ to
$\left|c\right\rangle$,and $\left|b\right\rangle$ to
$\left|c\right\rangle$, respectively. Meanwhile,
$\gamma_{12}=P\sqrt{\gamma_1\gamma_2}$ represents the spontaneously
generated interference which is resulted from the cross coupling
between the transitions
$\left|a\right\rangle\leftrightarrow\left|c\right\rangle$ and
$\left|b\right\rangle\leftrightarrow\left|c\right\rangle$, and
$P=\overrightarrow{b_{ac}}\cdot\overrightarrow{b_{bc}}/(\left|\overrightarrow{b_{ac}}\right|\cdot\left|\overrightarrow{b_{bc}}\right|)=\cos\theta$.
Here $\overrightarrow{b_{ac}}$ and $\overrightarrow{b_{bc}}$
represent the atomic dipole polarizations and $\theta$ is the angle
between the two dipole moments. From the expressions of the
parameter P and $\gamma_{12}$ we can find that the spontaneously
generated interference dependents on the angle between the two
dipole moments. When the two dipole moments are perpendicular to
each other the interference effect disappears (p=0), and it will be
maximal (p=1) if the two dipole moments are parallel to each other.
Note that $\kappa_j$ ($j=1,2$) are the damping constants of two
cavity modes

Based on the standard methods of laser theory in \cite{39},
considering the spontaneously decay of atom, $\rho_{ac}$ and
$\rho_{bc}$ can be evaluated to the first order in the coupling
constants $g_1$ and $g_2$ as
\begin{eqnarray}
\label{eq5a}
&&i\dot{\rho}_{ac}=-i\gamma_1\rho_{ac}-i\gamma_{12}\rho_{bc}+\Delta\rho_{ac}-\Omega\rho_{bc}-g_1\rho_{aa}^{(0)}a_1-g_2\rho_{ab}^{(0)}a_2+g_1a_1\rho_{cc}^{(0)},\\
\label{eq5b}
&&i\dot{\rho}_{bc}=-i\gamma_2\rho_{bc}-i\gamma_{12}\rho_{ac}+\Delta\rho_{bc}-\Omega^{*}\rho_{ac}-g_1\rho_{ba}^{(0)}a_1-g_2\rho_{bb}^{(0)}a_2+g_2a_2\rho_{cc}^{(0)},
\end{eqnarray}
\noindent where the density matrix elements $\rho_{ij}^{(0)}$ can be
obtained by the corresponding zeroth-order equations
\begin{eqnarray}
\label{eq6a}
&&i\dot{\rho}_{aa}^{(0)}=-2i\gamma_1\rho_{aa}^{(0)}-i\gamma_{12}(\rho_{ba}^{(0)}+\rho_{ab}^{(0)})-\Omega\rho_{ba}^{(0)}+\Omega^{*}\rho_{ab}^{(0)}+\Omega_a\rho_{f},\\
\label{eq6b}
&&i\dot{\rho}_{bb}^{(0)}=-2i\gamma_2\rho_{bb}^{(0)}-i\gamma_{12}(\rho_{ba}^{(0)}+\rho_{ab}^{(0)})-\Omega^{*}\rho_{ab}^{(0)}+\Omega\rho_{ba}^{(0)},\\
\label{eq6c}
&&i\dot{\rho}_{ab}^{(0)}=-i(\gamma_1+\gamma_2)\rho_{ab}^{(0)}-i\gamma_{12}(\rho_{aa}^{(0)}+\rho_{bb}^{(0)})-\Omega\rho_{bb}^{(0)}+\Omega\rho_{aa}^{(0)},\\
\label{eq6d}
&&i\dot{\rho}_{ba}^{(0)}=-i(\gamma_1+\gamma_2)\rho_{ba}^{(0)}-i\gamma_{12}(\rho_{aa}^{(0)}+\rho_{bb}^{(0)})+\Omega^{*}\rho_{bb}^{(0)}-\Omega^{*}\rho_{aa}^{(0)},\\
\label{eq6e} &&i\dot{\rho}_{cc}^{(0)}=0.
\end{eqnarray}
\noindent Plugging the steady-state solution of $\rho_{ij}^{(0)}$
into Eqs. (\ref{eq5a}-\ref{eq5b}), we find the steady-state solution
for $\rho_{ac}$ and $\rho_{bc}$ can be described as:
\begin{eqnarray}
\label{eq7a}
&&ig_1\rho_{ac}=\alpha_{11}\rho_fa_1+\alpha_{12}\rho_fa_2,\\
\label{eq7b}
&&ig_2\rho_{bc}=\alpha_{21}\rho_fa_1+\alpha_{22}\rho_fa_2,
\end{eqnarray}
\noindent with the explicit expressions of the coefficients
$\alpha_{ij}$ being given in Appendix. By substituting
Eqs.(\ref{eq7a}-\ref{eq7b}) to the Eq. (\ref{eq2}), the reduced
master equation govern the evolution of the cavity field can be
obtained as
\begin{eqnarray}
\label{eq8}
\dot{\rho}_{f}&=&-\alpha_{11}(a_1^{\dag}\rho_fa_1-\rho_fa_1a_1^{\dag})-\alpha_{12}(a_1^{\dag}\rho_fa_2-\rho_fa_2a_1^{\dag})\nonumber\\
&&-\alpha_{22}(a_2^{\dag}\rho_fa_2-\rho_fa_2a_2^{\dag})-\alpha_{21}(a_2^{\dag}\rho_fa_1-\rho_fa_1a_2^{\dag})\nonumber\\
&&+\kappa_{1}(a_1\rho_fa_1^{\dag}-\rho_fa_1^{\dag}a_1)+\kappa_{2}(a_2\rho_fa_2^{\dag}-\rho_fa_2^{\dag}a_2)+H.c..
\end{eqnarray}

Here we remain all orders in the Rabi frequency $\Omega$ whereas
only consider second order in the coupling constants $g_1$, $g_2$
due to that the coupling constants of two cavity modes are smaller
than other system parameters in our scheme. Thus we can ignore the
saturation effects and operate in the regime of linear
amplification.

\section{entanglement of the cavity fields}
In this section, we use the sufficient inseparability criterion
proposed by Duan et al \cite{28}. to verify that CV entanglement
with large mean number of photons can be obtained in our model, and
study the property of entanglement under the given conditions.

According to Duan's criterion \cite{28}, the two cavity modes are
entangled if and only if the sum of the variances of the two
Einstein-Podolsky-Rosen (EPR) type operators
$\hat{u}=\hat{x}_1+\hat{x}_2$ and $\hat{v}=\hat{p}_1+\hat{p}_2$
satisfies the following inequality
\begin{equation}
\label{eq9} \left\langle(\Delta \hat {u})^2+(\Delta \hat
{v})^2\right\rangle<2,
\end{equation}
\noindent with the pair quadrature operators $\hat
{x}_j=(a_j+a_j^{\dag})/\sqrt{2}$ and $\hat
{p}_j=-i(a_j-a_j^{\dag})/\sqrt{2}$ ($j=1,2$) the local operators
which correspond to the mode at the frequency $\nu_j$. By
substituting $\hat{x}_j$ and $\hat{p}_j$ into equation (\ref{eq9}),
we can express the total variance of the operators $\hat{u}$ and
$\hat{v}$ in terms of the operators $a_j$ and $a_j^{\dag}$ and
achieve
\begin{eqnarray}
\label{eq10} \left\langle(\Delta
\hat{u})^2+(\Delta\hat{v})^2\right\rangle &=& 2[1+\langle
a_1^{\dag}a_1\rangle+\langle a_2^{\dag}a_2\rangle+\left\langle
a_1a_2\right\rangle+\langle
a_1^{\dag}a_2^{\dag}\rangle\nonumber\\
&&-\left\langle a_1\right\rangle\langle
a_1^{\dag}\rangle-\left\langle a_2\right\rangle\langle
a_2^{\dag}\rangle-\left\langle a_1\right\rangle\left\langle
a_2\right\rangle-\langle a_1^{\dag}\rangle\langle
a_2^{\dag}\rangle].
\end{eqnarray}

With the help of equation (\ref{eq8}),we can obtain the equations of
motion for the expectation values of the field operators in equation
(\ref{eq10}) as
\begin{eqnarray}
\label{eq12a} &&\frac{\partial}{\partial t}\langle
a_1\rangle=-(\alpha_{11}+\kappa_1)\langle
a_1\rangle-\alpha_{12}\langle a_2\rangle,\\
\label{eq12b} &&\frac{\partial}{\partial t}\langle
a_2\rangle=-(\alpha_{22}+\kappa_2)\langle
a_2\rangle-\alpha_{21}\langle a_1\rangle,\\
\label{eq12c} &&\frac{\partial}{\partial t}\langle
a_1^{\dag}\rangle=-(\alpha_{11}^{*}+\kappa_1)\langle
a_1^{\dag}\rangle-\alpha_{12}^{*}\langle a_2^{\dag}\rangle,\\
\label{eq12d} &&\frac{\partial}{\partial t}\langle
a_2^{\dag}\rangle=-(\alpha_{22}^{*}+\kappa_2)\langle
a_2^{\dag}\rangle-\alpha_{21}^{*}\langle a_1^{\dag}\rangle,\\
\label{eq12e} &&\frac{\partial}{\partial t}\langle
a_1^{\dag}a_1\rangle=-(\alpha_{11}+\alpha_{11}^{*}+2\kappa_1)\langle
a_1^{\dag}a_1\rangle-\alpha_{12}\langle
a_1^{\dag}a_2\rangle-\alpha_{12}^{*}\langle
a_1a_2^{\dag}\rangle-(\alpha_{11}+\alpha_{11}^{*}),\\
\label{eq12f} &&\frac{\partial}{\partial t}\langle
a_2^{\dag}a_2\rangle=-(\alpha_{22}+\alpha_{22}^{*}+2\kappa_2)\langle
a_2^{\dag}a_2\rangle-\alpha_{21}^{*}\langle
a_1^{\dag}a_2\rangle-\alpha_{21}\langle
a_1a_2^{\dag}\rangle-(\alpha_{22}+\alpha_{22}^{*}),\\
\label{eq12g} &&\frac{\partial}{\partial t}\langle
a_1^{\dag}a_2\rangle=-\alpha_{21}\langle
a_1^{\dag}a_1\rangle-\alpha_{12}^{*}\langle
a_2^{\dag}a_2\rangle-(\alpha_{11}^{*}+\alpha_{22}+\kappa_1+\kappa_2)\langle
a_1^{\dag}a_2\rangle-(\alpha_{12}^{*}+\alpha_{21}),\\
\label{eq12h} &&\frac{\partial}{\partial t}\langle
a_1a_2^{\dag}\rangle=-\alpha_{21}^{*}\langle
a_1^{\dag}a_1\rangle-\alpha_{12}\langle
a_2^{\dag}a_2\rangle-(\alpha_{11}+\alpha_{22}^{*}+\kappa_1+\kappa_2)\langle
a_1a_2^{\dag}\rangle-(\alpha_{12}+\alpha_{21}^{*}),\\
\label{eq12i} &&\frac{\partial}{\partial t}\langle
a_1a_2\rangle=-\alpha_{21}\langle a_1a_1\rangle-\alpha_{12}\langle
a_2a_2\rangle-(\alpha_{11}+\alpha_{22}+\kappa_1+\kappa_2)\langle
a_1a_2\rangle,\\
\label{eq12j} &&\frac{\partial}{\partial t}\langle
a_1a_1\rangle=-2(\alpha_{11}+\kappa_1)\langle
a_1a_1\rangle-2\alpha_{12}\langle
a_1a_2\rangle,\\
\label{eq12k} &&\frac{\partial}{\partial t}\langle
a_2a_2\rangle=-2(\alpha_{22}+\kappa_2)\langle
a_2a_2\rangle-2\alpha_{21}\langle
a_1a_2\rangle,\\
\label{eq12l} &&\frac{\partial}{\partial t}\langle
a_1^{\dag}a_2^{\dag}\rangle=-\alpha_{21}^{*}\langle
a_1^{\dag}a_1^{\dag}\rangle-\alpha_{12}^{*}\langle
a_2^{\dag}a_2^{\dag}\rangle-(\alpha_{11}^{*}+\alpha_{22}^{*}+\kappa_1+\kappa_2)\langle
a_1^{\dag}a_2^{\dag}\rangle,\\
\label{eq12m} &&\frac{\partial}{\partial t}\langle
a_1^{\dag}a_1^{\dag}\rangle=-2(\alpha_{11}^{*}+\kappa_1)\langle
a_1^{\dag}a_1^{\dag}\rangle-2\alpha_{12}^{*}\langle
a_1^{\dag}a_2^{\dag}\rangle,\\
\label{eq12n} &&\frac{\partial}{\partial t}\langle
a_2^{\dag}a_2^{\dag}\rangle=-2(\alpha_{22}^{*}+\kappa_2)\langle
a_2^{\dag}a_2^{\dag}\rangle-2\alpha_{21}^{*}\langle
a_1^{\dag}a_2^{\dag}\rangle.
\end{eqnarray}

By numerically solving these equations, we can give a few numerical
results for the time evolution of the total photon numbers
$\langle\hat{N}\rangle=\langle a_1^{\dag}a_1\rangle+\langle
a_2^{\dag}a_2\rangle$ and
$\left\langle(\Delta\hat{u})^2+(\Delta\hat{v})^2\right\rangle$ with
different values of parameter while the cavity modes are assumed to
be in the coherent state $\left|10,-10\right\rangle$, as illustrated
in Figs. 2-4.  It is easy to find that, the CV entanglement with
large mean number of photons  from system of V-type atom in a
quantum beat laser can be generated, and the entanglement can be
enhanced by the spontaneously generated interference. For
simplicity, all the parameters used here are scaled with $g$, and we
have chosen $\phi=\pi/2$ during our numerical calculations.

By above knowable, $\gamma_{12}=P\sqrt{\gamma_1\gamma_2}$ represents
the spontaneously generated interference. Therefore, in order to
check the effect of the spontaneously generated interference on the
time evolution of entanglement, we can study the property of
entanglement for different values of atomic decay rates and the
parameter P.

In Fig. 2, we show the plot of the time development of
$\langle(\Delta\hat{u})^2+(\Delta\hat{v})^2\rangle$ and
$\langle\hat{N}\rangle$ for different values of atomic decay rates,
when the cavity field is initially in the coherent state
$\left|10,-10\right\rangle$. From Fig. 2(a), one can find that the
intensity and period of entanglement can be enlarged by increasing
spontaneous emission decay rates of the atom level, while Fig. 2(b)
illustrates that the effect of the atomic decay rates $\gamma_j$
($j=1,2$) on the maximum mean photon numbers give a small extent.

\begin{figure}\label{fig2}
\includegraphics[width=7cm]{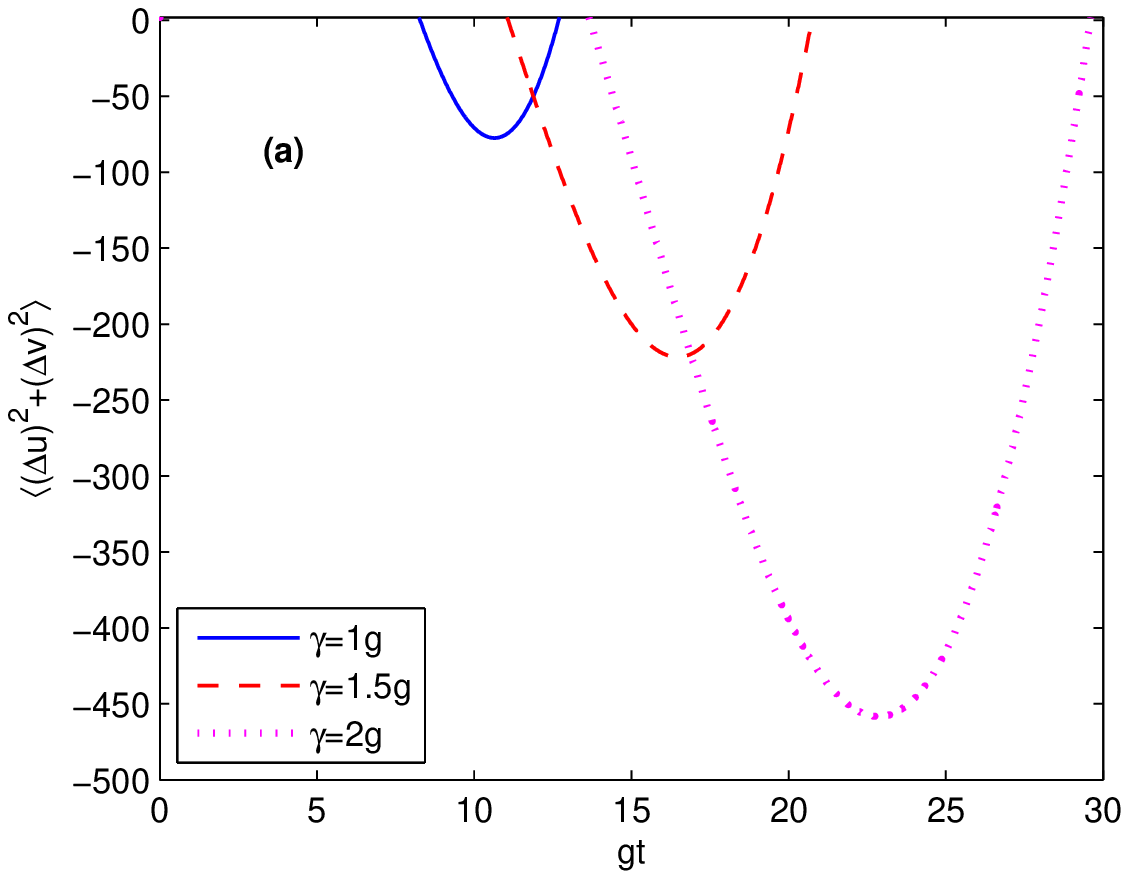}
\includegraphics[width=7cm]{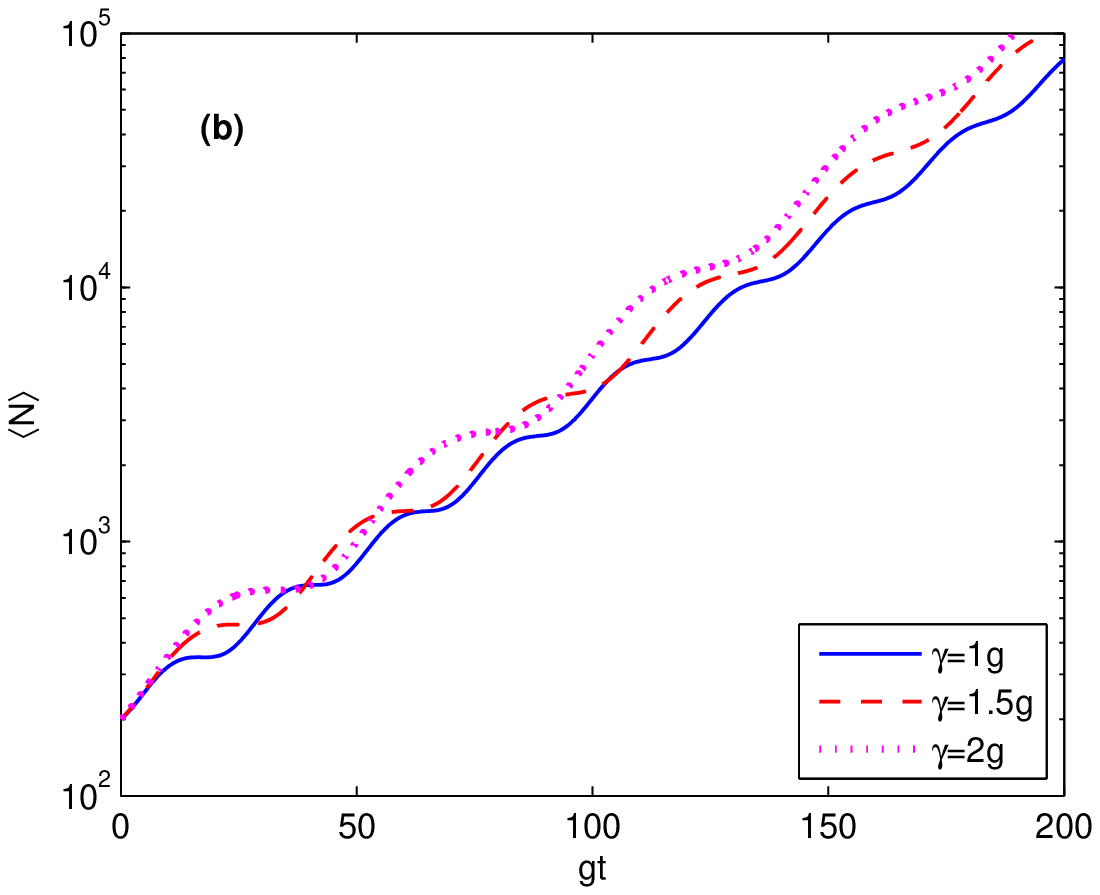}
\caption{\label{fig2} The time evolution of
$\langle(\Delta\hat{u})^2+(\Delta\hat{v})^2\rangle$ (shown in Fig.
2(a)) and the total mean photon numbers $\langle\hat{N}\rangle$
(shown in Fig. 2(b)) for different atomic decay rates $\gamma$
($\gamma_1=\gamma_2=\gamma$), when the cavity field is initially in
the coherent state $\left|10,-10\right\rangle$. The other parameters
are $\kappa_1=\kappa_2=0.001g$, $P=0.5$, $|\Omega|=10g$,
$\Delta_1=\Delta_2=\Delta=g$, $\gamma_a=5g$ and $\phi=\pi/2$.}
\end{figure}
\begin{figure}\label{fig3}
\includegraphics[width=7cm]{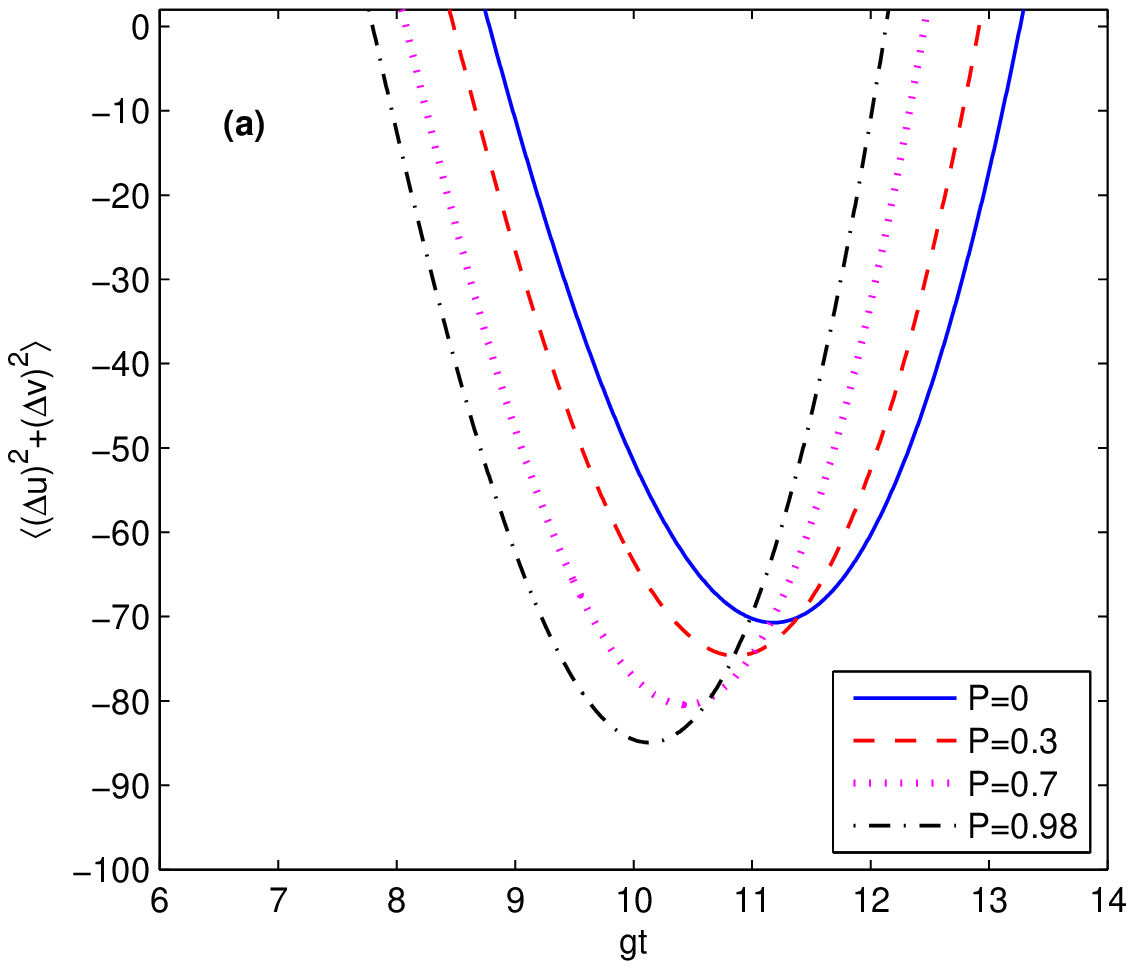}
\includegraphics[width=7cm]{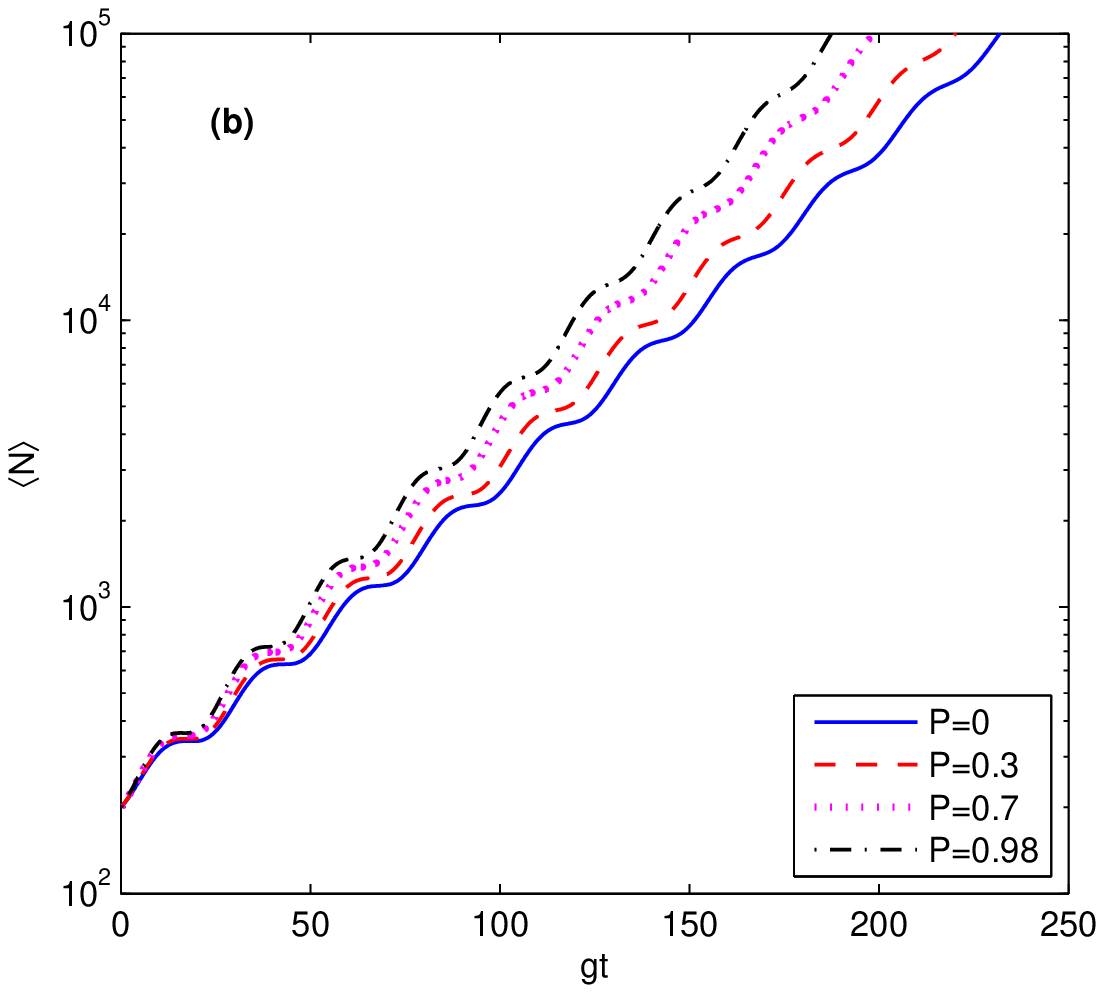}
\caption{\label{fig3} The time evolution of
$\langle(\Delta\hat{u})^2+(\Delta\hat{v})^2\rangle$ (shown in Fig.
3(a)) and the total mean photon numbers $\langle\hat{N}\rangle$
(shown in Fig. 3(b)) different values of P, when the cavity field is
initially in the coherent state $\left|10,-10\right\rangle$. The
other parameters are $\kappa_1=\kappa_2=0.001g$,
$\gamma_1=\gamma_2=\gamma=g$, $|\Omega|=10g$,
$\Delta_1=\Delta_2=\Delta=g$, $\gamma_a=5g$ and $\phi=\pi/2$.}
\end{figure}

In order to check the influence of P on the entanglement property,
we numerically simulate the time evolution of
$\langle(\Delta\hat{u})^2+(\Delta\hat{v})^2\rangle$ and
$\langle\hat{N}\rangle$ for different values of P. As shown in Fig.
3(a), when  the cavity field is initially in coherent state
$\left|10,-10\right\rangle$, the intensity of entanglement between
the two cavity modes slightly enhances with the increase of the
value of P. With the same set of the parameters, Fig. 3(b) shows
that the maximum mean photon numbers $\langle\hat{N}\rangle$ become
more pronounced with the increase of the value of P.

Up to now, we have investigated the the influence of spontaneous
emission decay rates of the atom level and the parameter P on the
time evolution of entanglement. It is easy to find that, the
entanglement can be enhanced by no matter increasing the spontaneous
emission decay rates of the atom level or increasing the value of P.
Then we conclude that, the spontaneously generated interference
which depends on $\gamma_j$ ($j=1,2$) and P, can strengthen the
entanglement.

We plot the influence of frequency detuning of the cavity field on
the time evolution of entanglement under condition that the cavity
field is initially in the coherent state $\left|10,-10\right\rangle$
(shown in Fig. 4). As shown in Fig. 4(a) that the intensity and
period of entanglement between the two cavity modes can be enlarged
at one time by decreasing the frequency detuning of the pump field
$\Delta$($\Delta_1=\Delta_2=\Delta$). In addition, Fig. 4(b)
illustrate that with the increased of the detuning $\Delta$, the
maximum mean photon number is enlarged. This result implicates that
in order to obtain the entanglement of cavity modes with high
intensity and longer period we can do that by properly adjusting the
frequency detuning.

\begin{figure}\label{fig4}
\includegraphics[width=7cm]{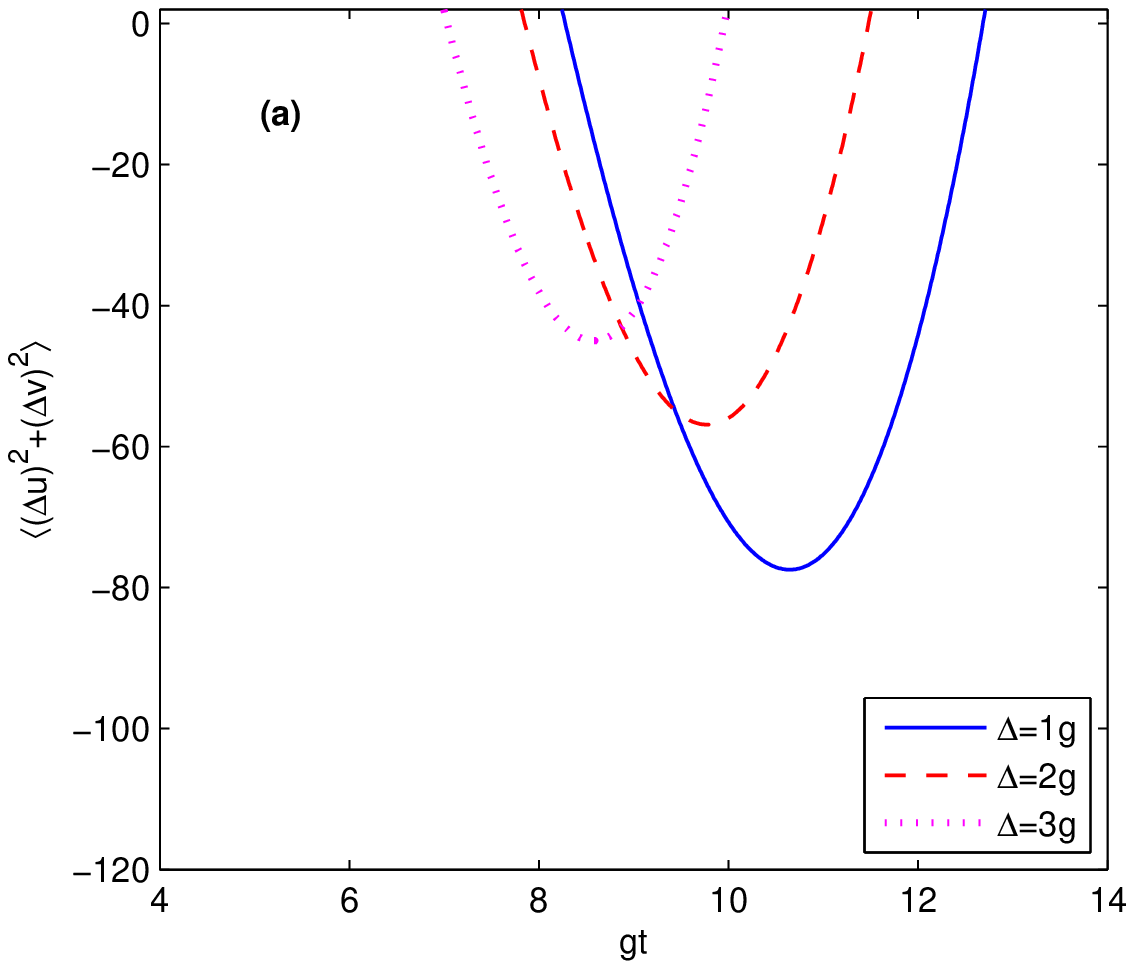}
\includegraphics[width=7cm]{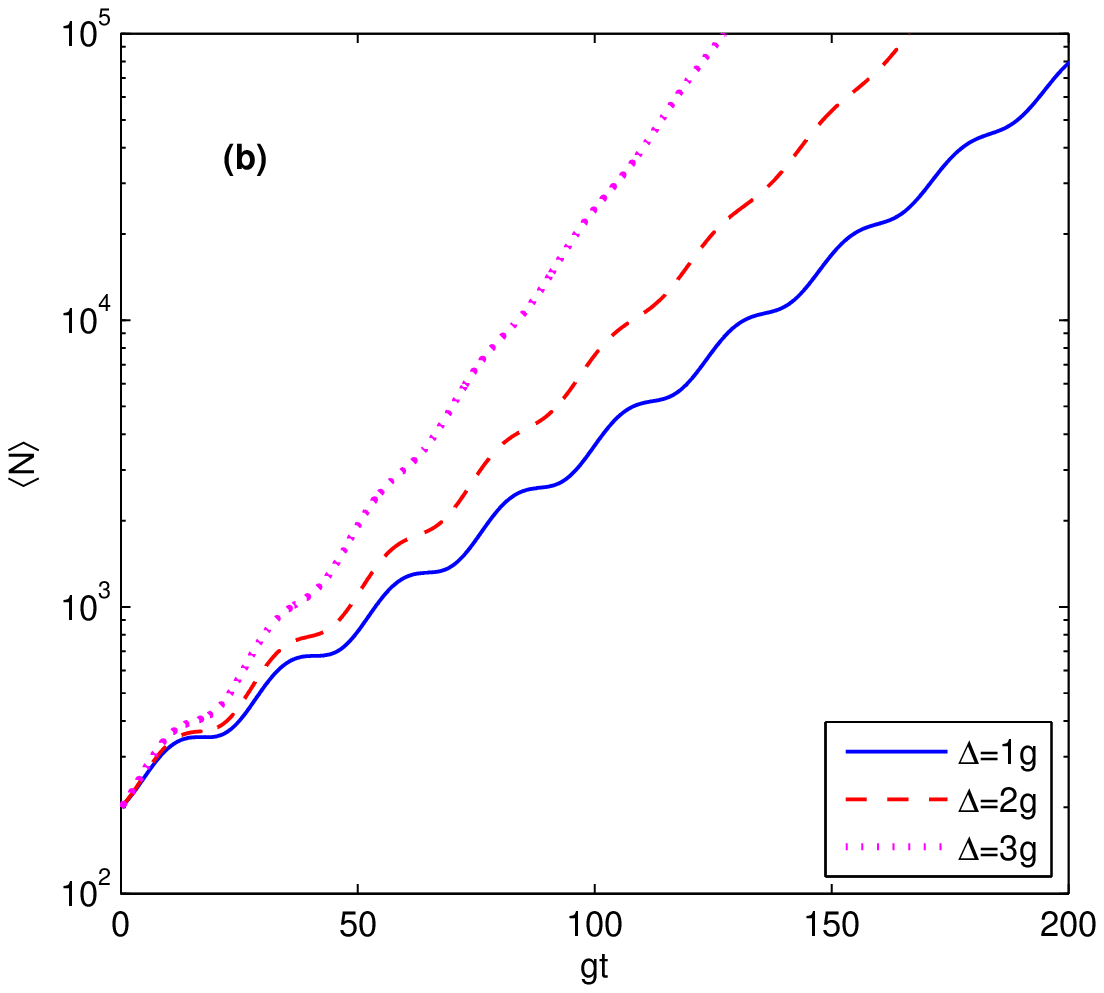}
\caption{\label{fig4} The time evolution of
$\langle(\Delta\hat{u})^2+(\Delta\hat{v})^2\rangle$ (shown in
Fig.4(a)) and the total mean photon numbers $\langle\hat{N}\rangle$
(shown in Fig. 4(b)) for different frequency detuning
$\Delta$($\Delta_1=\Delta_2=\Delta$), when the cavity field is
initially in the coherent state $\left|10,-10\right\rangle$. The
other parameters are $\gamma_1=\gamma_2=g$,
$\kappa_1=\kappa_2=0.001g$, $P=0.5$, $|\Omega|=10g$, $\gamma_a=5g$
and $\phi=\pi/2$.}
\end{figure}

Before conclusion, we should note that our scheme is drastically
different from the conventional scheme of CV entanglement generation
\cite{32,34}. In our scheme, with decreasing the frequency detuning,
our numerical results showed that a long entanglement time and
strong entanglement intensity can be synchronously achieved. It
illustrates that the entanglement of cavity modes with higher
intensity and longer period can be realized in our scheme with a
low-Q cavity. And it can strengthen the entanglement by increasing
the spontaneous emission decay rates of the atom level and the value
of the parameter P. The physical reason can be explained that no
matter larger the atom decay rates $\gamma_j$ ($j=1,2$) or larger
the value of P results in increasing the spontaneously generated
interference which can enhance the entanglement. All these
distinguish advances illustrate that our scheme is drastically
different from the conventional scheme.

\section{conclusion}

In summary, we have proposed a new scheme to generate the CV
entanglement and investigated the evolution of it from a system of
V-type atom in a quantum beat laser \cite{33}. In this scheme, the
two transitions in the V-type atom independently interact with the
two cavity modes while the two upper levels of the atom are driven
by a strong classical field. By taking into account the effects of
spontaneously generated interference between two atomic decay
channels, and using the standard methods of laser theory \cite{39},
we show that, in the given conditions, the CV entanglement with
large mean number of photons can be realized in our scheme. And by
properly modulating the frequency detuning of the field can adjust
the entanglement period, intensity and the total mean photon numbers
of two cavity modes. Different from the conventional scheme
\cite{32,34}, the CV entanglement of cavity modes with higher
intensity and longer period can be realized in our scheme with a
low-Q cavity. Furthermore, our results showed that the entanglement
can be significantly enhanced by the spontaneously generated
interference.

We would like to thank Prof. Peng Xue for her enlightening discussions. The research is supported in part by National Natural Science Foundation of China under Grant Nos. 11374050 and 61372102, by Qing Lan project of Jiangsu, and by the Fundamental Research Funds for the Central Universities under Grant No. 2242012R30011.

\appendix

\section{coefficients}

Here we give the expressions of the coefficients $A_{ij}$ and
$B_{ij}$ ($i,j=1,2$) in Eqs. (\ref{eq7a}-\ref{eq8}),

\begin{eqnarray}
\label{eqA1}
&&\alpha_{11}=-g_1^2[(R+i\Delta)\beta_{aa}+(i\Omega-\gamma_{12})\beta_{ba}]/D_1,\\
\label{eqA2}
&&\alpha_{12}=-g_1g_2[(R+i\Delta)\beta_{ab}+(i\Omega-\gamma_{12})\beta_{bb}]/D_1,\\
\label{eqA3}
&&\alpha_{22}=-g_2^2[(R+i\Delta)\beta_{bb}+(i\Omega^*-\gamma_{12})\beta_{ab}]/D_1,\\
\label{eqA4}
&&\alpha_{21}=-g_1g_2[(R+i\Delta)\beta_{ba}+(i\Omega^*-\gamma_{12})\beta_{aa}]/D_1,\\
\label{eqA5}
&&\beta_{aa}=(2R^2-\gamma_{12}^2+|\Omega|^2)R\gamma_a/D_2,\\
\label{eqA6}
&&\beta_{bb}=(\gamma_{12}+i\Omega)(\gamma_{12}-i\Omega^*)R\gamma_a/D_2,\\
\label{eqA7}
&&\beta_{ab}=-(\gamma_{12}+i\Omega)(2R^2-i\gamma_{12}\Omega-i\gamma_{12}\Omega^*)R\gamma_a/(2D_2),\\
\label{eqA8}
&&\beta_{ba}=-(\gamma_{12}-i\Omega^*)(2R^2+i\gamma_{12}\Omega+i\gamma_{12}\Omega^*)R\gamma_a/(2D_2),
\end{eqnarray}
\noindent with
\begin{eqnarray}
\label{eqA9}
D_1&=&(R+i\Delta)^2-(i\Omega-\gamma_{12})(i\Omega^*-\gamma_{12}),\\
\label{eqA10}
D_2&=&4R^4+4R^2|\Omega|^2-4R^2\gamma_{12}^2-\gamma_{12}^2(\Omega+\Omega^*)^2,
\end{eqnarray}
where we have assumed $\gamma_1=\gamma_2=\gamma=R$.

\end{document}